# Comparison of highly-compressed *C2/m*-SnH$_{12}$ superhydride with conventional superconductors

E. F. Talantsev[1,2*]

[1]M.N. Mikheev Institute of Metal Physics, Ural Branch, Russian Academy of Sciences, 18, S. Kovalevskoy St., Ekaterinburg, 620108, Russia

[2]NANOTECH Centre, Ural Federal University, 19 Mira St., Ekaterinburg, 620002, Russia

*E-mail: evgeny.talantsev@imp.uran.ru

**Abstract**

Satterthwaite and Toepke (1970 *Phys. Rev. Lett.* **25** 741) predicted high-temperature superconductivity in hydrogen-rich metallic alloys, based on an idea that these compounds should exhibit high Debye frequency of the proton lattice, which boosts the superconducting transition temperature, $T_c$. The idea has got full confirmation more than four decades later when Drozdov *et al* (2015 *Nature* **525** 73) experimentally discovered near-room-temperature superconductivity in highly-compressed sulphur superhydride, H$_3$S. To date, more than a dozen of high-temperature hydrogen-rich superconducting phases in Ba-H, Pr-H, P-H, Pt-H, Ce-H, Th-H, S-H, Y-H, La-H, and (La,Y)-H systems have been synthesized and, recently, Hong *et al* (2021 *arXiv*:2101.02846) reported on the discovery of *C2/m*-SnH$_{12}$ phase with superconducting transition temperature of $T_c \sim 70$ K. Here we analyse the magnetoresistance data, $R(T,B)$, of *C2/m*-SnH$_{12}$ phase and report that this superhydride exhibits the ground state superconducting gap of $\Delta(0) = 9.2 \pm 0.5$ meV, the ratio of $2\Delta(0)/k_B T_c = 3.3 \pm 0.2$, and $0.010 < T_c/T_F < 0.014$ (where $T_F$ is the Fermi temperature) and, thus, *C2/m*-SnH$_{12}$ falls into unconventional superconductors band in the Uemura plot.

# Comparison of highly-compressed $C2/m$-SnH$_{12}$ superhydride with conventional superconductors

## I. Introduction

Satterthwaite and Toepke [1] were first who understood that hydrogen-rich compound should exhibit highest superconducting transition temperature: "…There has been theoretical speculation [2] that metallic hydrogen might be a high-temperature superconductor, in part because of the very high Debye frequency of the proton lattice. With high concentrations of hydrogen in the metal hydrides one would expect lattice modes of high frequency and if there exists an attractive pairing interaction one might expect to find high-temperature superconductivity in these systems also." Mathematical rigorous description of Satterthwaite's and Toepke's idea [1] had been given 34 years later by Ashcroft [3].

In 2015 Drozdov *et al* [4] reported on experimental discovery of first near-room-temperature superconductor (NRTS) H$_3$S, which was also the first superhydride compound synthesized at megabar pressure level heated by laser pulses inside of diamond anvil cell. This technique is used since than to synthesize new NRTS phases, and to date more than a dozen high-temperature hydrogen-rich superconducting phases have been synthesised in Pr-H [5], Ba-H [6], P-H [7], Pt-H [8], Ce-H [9], Th-H [10,11], S-H [4,12-17], Y-H [18,19], La-H [20-24], (La,Y)-H [25] and CaH$_x$ [26,27] systems.

Recently, Hong *et al* [28] extended superhydride family by the discovery of $C2/m$-SnH$_{12}$ phase which exhibits the superconducting transition temperature of $T_c \sim 70$ K at pressure of $P = 190$ GPa. This experimental result is in a good accord with first-principles calculations performed in 2015 by Esfahani *et al* [29], who predicted $T_c = 83$-$93$ K for $C2/m$-SnH$_{12}$ phase compressed at pressure of $P = 250$ GPa. Despite Esfahani *et al* [29] predicted that $C2/m$-SnH$_{12}$ phase can be thermodynamically stable at $P \geq 250$ GPa, XRD studies [28] show that $C2/m$-SnH$_{12}$ phase is dominant at lower pressure range of $P \sim 200$ GPa. This difference can be explained by an atomic disorder, hydrogen non-stoichiometry, etc., which are always (in some degree) real world samples features. It should be noted, that here we assume that calculated values for the electron-phonon coupling constant, $\lambda_{e\text{-}ph} = 1.25$, and logarithmic average phonon frequency, $\hbar \cdot \omega_{log} = 991\ K$, reported by Esfahani *et al* [29] for $C2/m$-$SnH_{12}$ compressed at $P = 250$ GPa will be still valid for sample compressed at $P = 190$ GPa [28].

Hong *et al* [28] measured magnetoresistance curves, $R(T,B)$, up to applied magnetic field of $B_{appl} = 7$ T, from which, by applying analytical equation proposed by Jones *et al* [30]:

$$B_{c2}(T) = \frac{\phi_0}{2\cdot\pi\cdot\xi^2(0)} \cdot \left(\frac{1-\left(\frac{T}{T_c}\right)^2}{1+\left(\frac{T}{T_c}\right)^2}\right) \tag{1}$$

where $\phi_0 = \frac{h}{2\cdot e}$ is superconducting flux quantum, and $\xi(0)$ is the ground state coherence length, the ground state upper critical field was deduced as $B_{c2}(0) = 11.2$ T.

Here we perform further analysis of $R(T,B)$ data reported by Hong *et al* [28] with the purpose to extract the ground state amplitude of the superconducting energy gap, $\Delta(0)$, one of primary parameters of the superconducting state. In addition, we calculate the ratio of transition temperature to the Fermi temperature, $T_F$, to locate $C2/m$-$SnH_{12}$ phase in Uemura plot [31,32].

**II. $R(T,B)$ analysis**

Primary task in the analysis of $R(T,B)$ data is to deduce the superconducting critical temperature, $T_c$, for which we recently proposed [33] to use a fit of experimental $R(T,B)$ data to a function:

$$R(T, B_{appl}) = R_0 + k \cdot T + \theta(T_c^{onset} - T) \cdot \left(\frac{R_{norm}}{\left(I_0\left(F\cdot\left(1-\frac{T}{T_c^{onset}}\right)^{3/2}\right)\right)^2}\right) + $$

$$\theta(T - T_c^{onset}) \cdot (R_{norm} + (k - k_1) \cdot T_c^{onset} + k_1 \cdot T) \quad (2)$$

where $R_0, R_{norm}, T_c^{onset}, k, k_1,$ and $F$ are free-fitting parameters, and $\theta(x)$ is the Heaviside function.

The first two terms in the Eq. 2, i.e. $(R_0 + k \cdot T)$, are introduced in Ref. 33 to adopt possible ohmic resistance in $R(T,B)$ curve which appears as a result of metallic weak-links in NRTS sample in diamond anvil cell.

The third fitting term in Eq. 2 which approximates the superconducting transition:

$$R(T, B_{appl}) = \frac{R_{norm}}{\left(I_0\left(F \cdot \left(1 - \frac{T}{T_c^{onset}}\right)^{3/2}\right)\right)^2} \quad (3)$$

was proposed by Tinkham [34] to fit experimental $R(T,B)$ curves in HTS cuprate ceramics, where Tinkham [34] proposed to use:

$$F = \frac{C}{2 \cdot B_{appl}} \quad (4)$$

where $C$ is free-fitting parameter having unit of Tesla, and $B_{appl}$ is applied magnetic field.

Physical background of Eq. 3 was explained by Tinkham [34] as: " *... the specific predicted $B^{3/2}$ dependence fits quite well with a variety of published data .... We also point out that the result ... would hold even if the functional form* (which is in our case Eqs. 3,4) *were replaced by some other similar function of $U_0/k_BT$, so long as the form of* (which is our Eq. 7) *holds."*

In this explanation, Tinkham [34] mentioned the ratio $U_0/k_BT$, where $k_B$ is the Boltzmann constant, and $U_0$ is a magnetic flux creep activation energy:

$$U_0 = \beta \cdot B_c^2 \frac{\phi_0 \cdot \xi}{\mu_0 \cdot B_{appl}} \quad (5)$$

where, $\beta$ is (presumed ~1) a constant which absorbed all numerical factors, , $\xi$ is superconducting coherence length, $B_{appl}$ is applied magnetic field, and $B_c$ is the thermodynamic field:

$$B_c = \frac{\phi_0}{2\cdot\sqrt{2}\cdot\pi\cdot\lambda\cdot\xi} \tag{6}$$

where λ is the London penetration depth. After further consideration, Tinkham [34] reported, that:

$$\frac{U_0}{k_B \cdot T} = \frac{A}{B_{appl}} \cdot \left(1 - \frac{T}{T_c^{onset}}\right)^{3/2} \tag{7}$$

where *A* is a constant of Tesla unit. Thus, in overall, Eq. 3 can be considered as a good approximation for the Abrikosov vortex flux creep. However, as it is mentioned by Tinkham [34], there are no restrictions to use other fitting functions which approximate $U_0/k_B T$ term in given superconductor.

As we discussed in previous paper [31], there is a significant disadvantage of Eq. 7, which remains in recent proposal for parameter *F* given by Hirsch and Marsiglio [35]:

$$F = \frac{1}{2\cdot\frac{B_{appl}}{B_{c2}(0)}} \tag{8}$$

that Eq. 3 cannot be used to fit $R(T, B_{appl} = 0)$ data, because the division by zero is prohibited. However, it was pointed out in Ref. 33, that there is no necessity for explicit use of $B_{appl}$ in the expression for parameter *F*, because $B_{appl}$ is known from experiment. Based on this, *F* can be free-fitting unitless value, which describe the sharpness of the transition.

However, it should be stressed that as it was mentioned by Tinkham [34] that: "…*some other similar function …*" can be used as well. And based on this, particular deduced *F* values are linked to main fitting term of $\left(I_0\left(F \cdot \left(1 - \frac{T}{T_c^{onset}}\right)^{3/2}\right)\right)^{-2}$ and as far as the goodness of fit is high, the fit will be in use to deduce $T_c^{onset}$ and $T_c$ within established strict mathematical routine, while particular *F* value has no practical use.

The fourth fitting term in Eq. 2, i.e. $(R_{norm} + (k - k_1) \cdot T_c^{onset} + k_1 \cdot T)$, represents a linear rise in the *R(T,B)* curve above the onset transition temperature, $T_c^{onset}$. More details about different terms in Eq. 2 can be found in Ref. 33.

Thus, if $R(T,B)$ fit to Eq. 2 has converged, $T_c$ can be defined at any $\frac{R(T)}{R(T_c^{onset})}$ criterion, for which in this work we used the $T_{c,0.05}$ criterion:

$$\frac{R(T)}{R(T_c^{onset})} = 0.05 \qquad (9)$$

Primary reasons why the superconducting critical temperature for highly-compressed superconductors should be defined at as low as practically possible $\frac{R(T)}{R(T_c^{onset})}$ ratio were discussed elsewhere [36]. Here we only point out that the use of $T_c^{onset}$ criterion, which utilizes in some, but not in all, reports on highly-compressed superconductors, can be objected by experimental fact that the change in $R(T)$ slope, or even sharp drop in $R(T)$, is observable at many phase transitions in condensed matter when structural phase transitions occur [37-39]. Classical example for this is the change in $R(T)$ slope at structural phase transitions α-γ and γ-ε in iron [40,41].

In addition to several fits for NRTS materials, which we showed in our previous work [33], in Fig. 1 we fit $R(T,B=0)$ data for $Fm\text{-}3m\text{-}LaH_{10}$ phase ($P$ = 138 GPa) for which experimental data has been recently reported by Sun *et al* [26]. The fit has high quality (with goodness of fit $R$ = 0.9981) and deduced $T_c^{onset}$ and $T_{c,0.05}$ are indicated in Fig. 1.

All fits presented in the manuscript have been performed by utilizing the Levenberg-Marquardt approach in non-linear fitting package of the Origin2017 software.

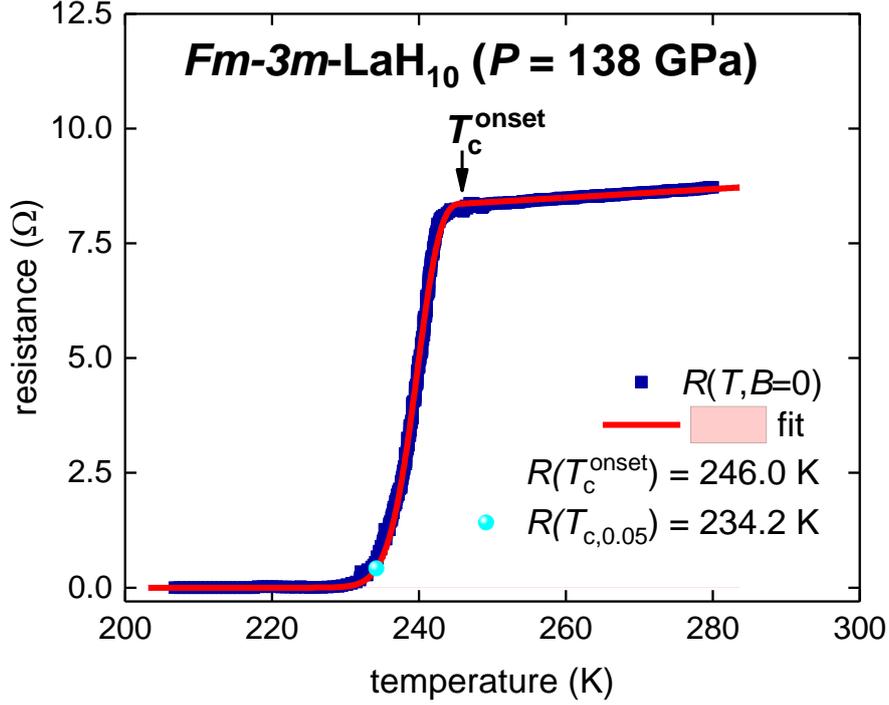

**Figure 1.** $R(T,B=0)$ data and fit to Eq. 1 for $Fm\text{-}3m\text{-}LaH_{10}$ ($P = 138$ GPa), where raw data was reported by Sun *et al* [24]. 95% confidence bars are shown by a pink shaded area; goodness of fit is $R = 0.9981$.

### III. Results

Fits to Eq. 2 of $R(T,B)$ data for $C2/m\text{-}SnH_{12}$ ($P = 190$ GPa) reported by Hong *et al* [28] are shown in Figs. 2,3, where Fig. 2 represents measurements performed at the "cooling" stage, while in Fig. 3 data and fits are shown for the "warming" stage. Despite a fact that $R(T,B)$ curves of $C2/m\text{-}SnH_{12}$ ($P = 190$ GPa) phase for "cooling" and "warming" stages are close to each other, these curves are not identical. For this reason, we deduce $T_{c,0.05}(B)$ for each stage with the purpose that full $B_{c2}(T)$ dataset will characterize as complete as practically possible the $C2/m\text{-}SnH_{12}$ phase. Results of the analysis are shown in Table 1.

It should be noted that $R(T,B)$ data for $C2/m\text{-}SnH_{12}$ ($P = 190$ GPa) reported by Hong *et al* [28] have linear ohmic term below transition temperature, which reflects the presence of metallic weak-links in the sample, which is accounted (as this mentioned above) by the term of $(R_0 + k \cdot T)$ in Eq. 2.

**Table 1.** Deduced $T_{c,0.05}(B)$ values for the "cooling" and the "warming" stages of *C2/m*-SnH$_{12}$ phase compressed at $P = 190$ GPa.

| Applied field, $B_{appl}$ (Tesla) | $T_{c,0.05}$ (cooling stage) (K) | $T_{c,0.05}$ (warming stage) (K) |
|---|---|---|
| 0 | 63.5 | 65.1 |
| 1 | 57.9 | 58.6 |
| 2 | 52.4 | 53.3 |
| 3 | 47.2 | 48.2 |
| 5 | 35.9 | 36.4 |
| 7 | 24.8 | 25.0 |

In overall, all fits have high-quality, even for $R(T,B=0)$ (Figs. 2,a and 3,a) for which the double transition is observed. For the latter the goodness of fit, $R = 0.9986$, while for the rest $R > 0.9989$.

It should be clarified, that as far as we have defined the critical temperature, $T_c$, by the $\frac{R(T)}{R(T_c^{onset})} = 0.05$ criterion (Eq. 9 and Table I), there is no any longer a need to write full designation, i.e. $T_{c,0.05}$, for this value because otherwise there will be a need to use the same subscript for other parameters, i.e. $B_{c2,0.05}(T)$, $\xi_{0.05}(0)$, $\Delta_{0.05}(0)$, etc.. Thus, in further analysis we omit the use of 0.05 designation in the subscripts, because when (which is implemented in many reports) $T_c$ and $B_{c2}(T)$ are defined by 50% of normal state resistance criterion, the designation of used criterion, i.e. $T_{c,0.50}$ and $B_{c2,0.50}(T)$, is always omitted (see, for instance, Ref. 28 where the latter criterion was used).

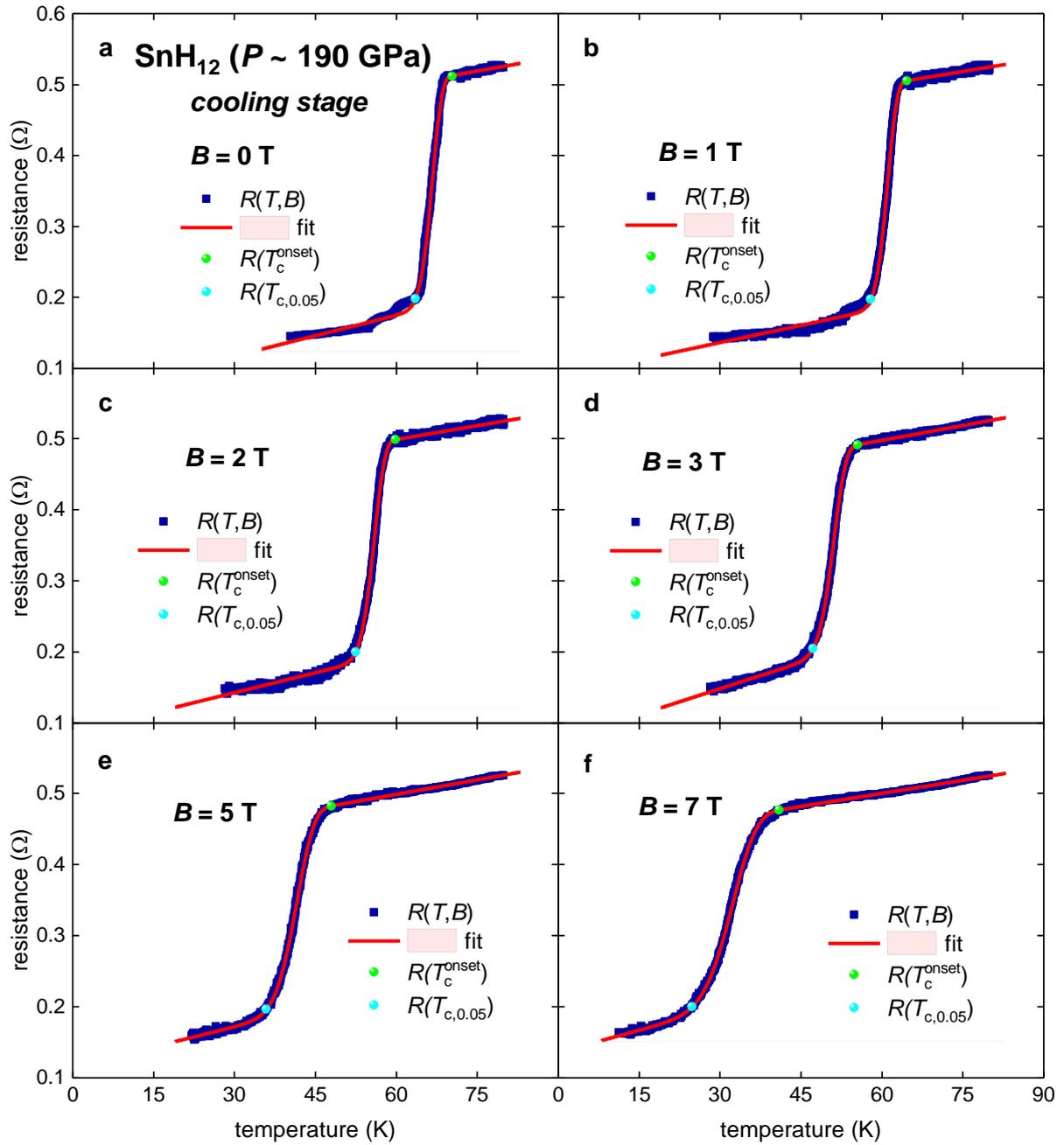

**Figure 2.** $R(T,B)$ data and fits to Eq. 1 for $C2/m$-SnH$_{12}$ ($P$ = 190 GPa) measured at *cooling* stage (raw data reported by Hong *et al* [26]). Goodness of fit is: (a) 0.9985, (b) 0.9990; (c) 0.9993; (d) 0.9996; (e) 0.9996; (f) 0.9996.

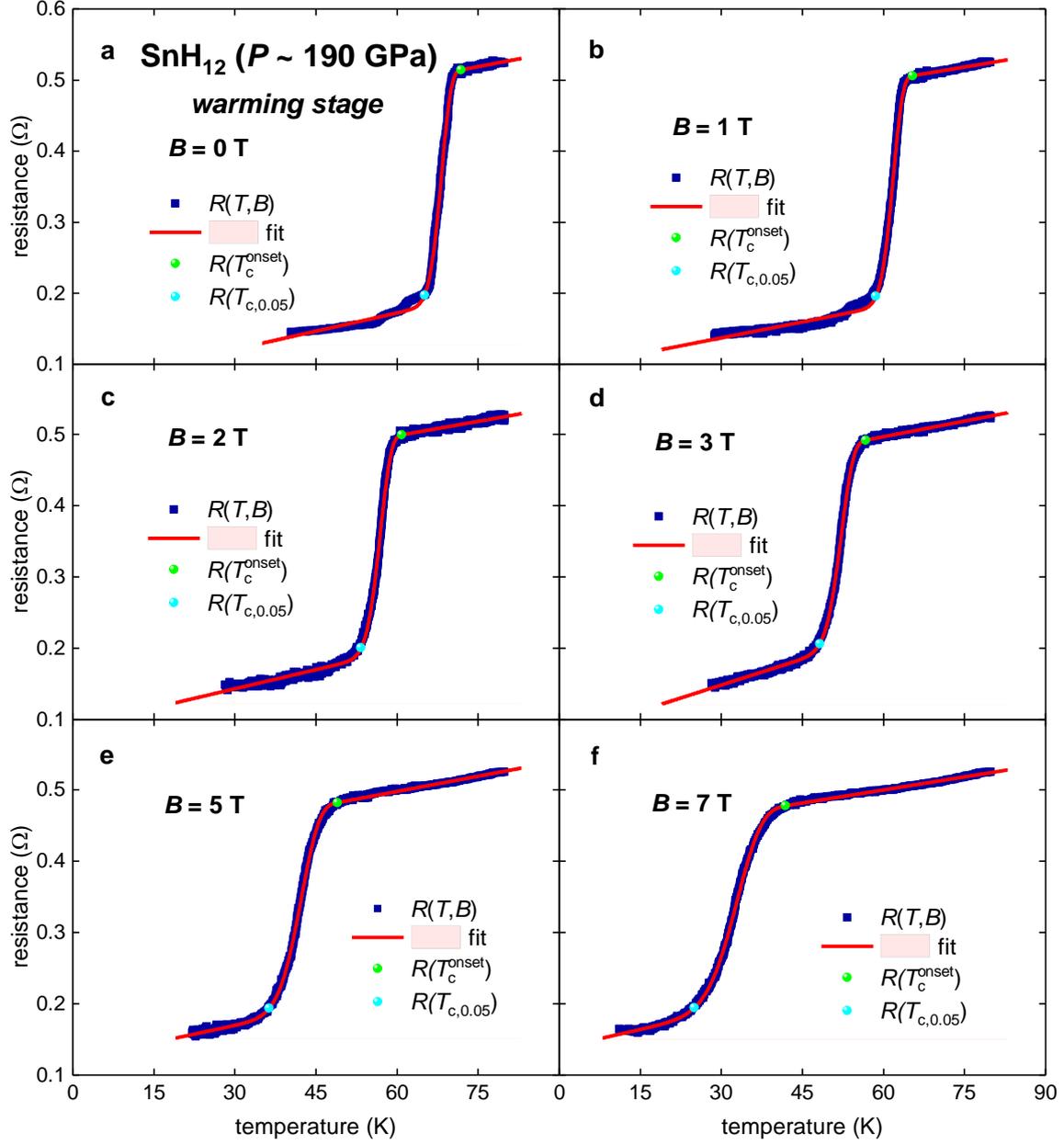

**Figure 3.** $R(T,B)$ data and fits to Eq. 1 for $C2/m$-SnH$_{12}$ ($P = 190$ GPa) measured at *warming* stage (raw data reported by Hong *et al* [28]). Goodness of fit is: (a) 0.9987, (b) 0.9990; (c) 0.9992; (d) 0.9996; (e) 0.9996; (f) 0.9996..

Deduced $T_c(B)$ values were used as raw $B_{c2}(T)$ data, which were fitted to upper critical field *s*-wave model [42]:

$$B_{c2}(T) = \frac{\phi_0}{2\cdot\pi\cdot\xi^2(0)} \cdot \left(\frac{1.77 - 0.43\cdot\left(\frac{T}{T_c}\right)^2 + 0.07\cdot\left(\frac{T}{T_c}\right)^4}{1.77}\right)^2 \cdot \left[1 - \frac{1}{2\cdot k_B\cdot T}\cdot\int_0^\infty \frac{d\varepsilon}{cosh^2\left(\frac{\sqrt{\varepsilon^2 + \Delta^2(T)}}{2\cdot k_B\cdot T}\right)}\right] \quad (10)$$

where $k_B$ is the Boltzmann constant, and the amplitude of temperature dependent superconducting gap, $\Delta(T)$, is given by [43,44]:

$$\Delta(T) = \Delta(0) \cdot \tanh\left[\frac{\pi \cdot k_B \cdot T_c}{\Delta(0)} \cdot \sqrt{\eta \cdot \frac{\Delta C}{C} \cdot \left(\frac{T_c}{T} - 1\right)}\right] \qquad (11)$$

where $\Delta C/C$ is the relative jump in electronic specific heat at $T_c$, and $\eta = 2/3$ for *s*-wave superconductors.

Eqs. 10,11 were used to extract $\xi(0)$, $\Delta(0)$, $T_c$ and $\frac{\Delta C}{C}$ in a variety of superconductors, for instance, in highly-compressed $H_3S$ [42], magic-angle twisted bilayer graphene [46], $V_3Si$ [47], $Nd_{0.8}Sr_{0.2}NiO_2$ [48] and iron-based superconductors [47]. Here we applied these equations to extract $\xi(0)$, $\Delta(0)$ and $\frac{2 \cdot \Delta(0)}{k_B \cdot T_c}$ in *C2/m*-$SnH_{12}$ (*P* = 190 GPa).

Eqs. 10,11 have four-free fitting parameters, $\xi(0)$, $\Delta(0)$, $T_c$, and $\Delta C/C$, i.e. the same number as one in the standard fitting function for the pinning force density, $F_p(B_{appl})$, [48-51]:

$$F_p(B_{appl}) = F_{p,max} \cdot \left(\frac{B_{appl}}{B_{c2}}\right)^p \cdot \left(1 - \frac{B_{appl}}{B_{c2}}\right)^q \qquad (12)$$

where $F_{p,max}$, $B_{c2}$, $p$ and $q$ are free-fitting parameters. Thus, Eqs. 9,10 can be characterized as a conventional mathematical tool in terms of the number of free-fitting parameters, where each deduced parameter has clear physical meaning.

It needs to be pointed out that *R(T,B)* curves were measured at only six $B_{appl}$ values, i.e. $B_{appl}$ = 0,1,2,3,5,7 T, which implies that conventional $B_{c2}(T)$ fit to Eqs. 10,11, where all four parameters are free, needs to be adopted for given $B_{c2}(T)$ dataset (it should be noted that usually [42,47] $B_{c2}(T)$ datasets have up to 30 raw upper critical field data). Thus, there is a need to reduce the number of free-fitting parameters in Eqs. 10,11. We used to fix $\frac{\Delta C}{C}$ value in our previous works [52-54] when experiments were performed over either a narrow temperature range, either at limited set of temperatures. Thus, we assumed that the relative jump in electronic specific heat at $T_c$ is equal to the Bardeen-Cooper-Schrieffer theory weak-coupling limit for *s*-wave superconductors [43,44,55,56]:

$$\frac{\Delta C}{C} = 1.43. \qquad (13)$$

That left in this case just $T_c$, $\xi(0)$ and $\Delta(0)$ as free fitting parameters in Eqs. 10,11. $B_{c2}(T)$ data fit to the restricted Eqs. 10,11 is shown in Fig. 4, where it can be seen that the fit has narrow 95% uncertainty bands and deduced parameters are $T_c = 64.6 \pm 0.3\ K$, $\xi(0) = 6.3 \pm 0.1\ nm$, $\Delta(0) = 9.15 \pm 0.51\ meV$, and

$$\frac{2 \cdot \Delta(0)}{k_B \cdot T_c} = 3.28 \pm 0.18. \qquad (14)$$

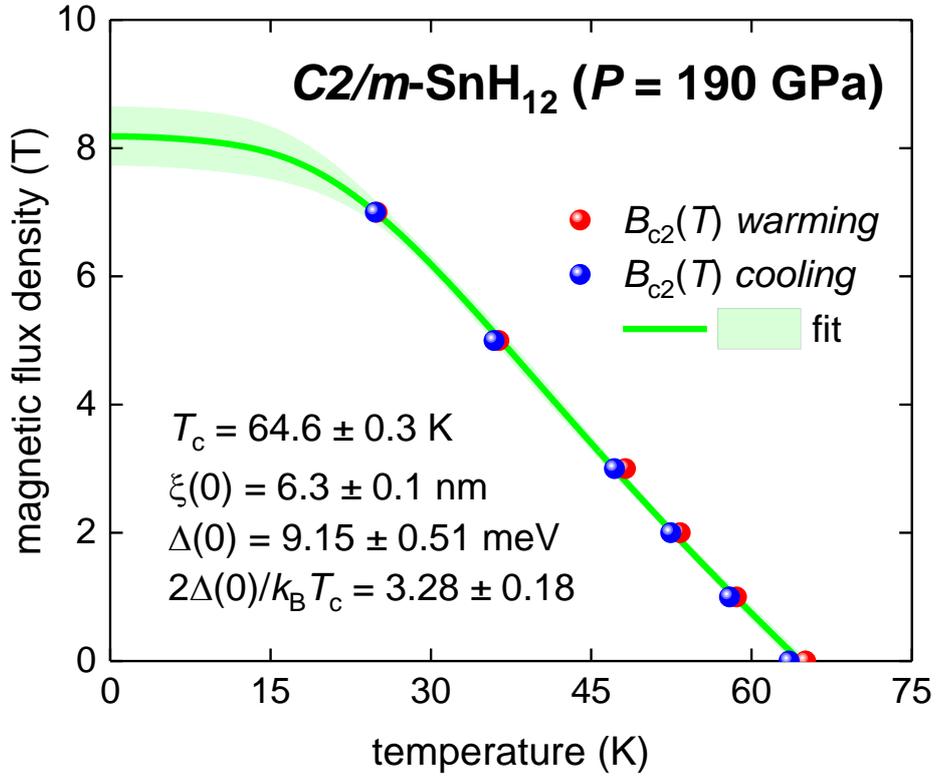

**Figure 4.** The upper critical field data, $B_{c2}(T)$, and data fit to Eqs. 3,4 for *C2/m*-SnH$_{12}$ (*P* = 190 GPa). $\frac{\Delta C}{C}$ was fixed to BCS weak-coupling limit of 1.43. 95% confidence bars are shown by a green shaded area; fit quality is $R = 0.9983$.

**IV. Comparison of *C2/m*-SnH$_{12}$ with conventional superconductors**

It might be appeared to be strange that deduced ratio of the gap amplitude to the transition temperature $\frac{2 \cdot \Delta(0)}{k_B \cdot T_C} = 3.28 \pm 0.18$ is lower than *s*-wave BCS weak coupling limit of [43,44,55,56]:

$$\frac{2 \cdot \Delta(0)}{k_B \cdot T_C} = 3.53 \tag{15}$$

However, if we assume that *C2/m*-SnH$_{12}$ (*P* = 190 GPa) has the Coulomb pseudopotential parameter, µ* = 0.13, which is weighted average value within many first principle calculations of NRTS materials (where µ* = 0.10-0.16 [5,6,9,10,18,25,29,57-71]), and, what is more important, that µ* = 0.13 was one of probable values used by Esfahani *et al* [29] in their predictive calculations for *C2/m*-SnH$_{12}$ phase, than the ratio of $\frac{k_B \cdot T_C}{\hbar \cdot \omega_{ln}}$ has got a value:

$$\frac{k_B \cdot T_C}{\hbar \cdot \omega_{ln}} = \frac{83}{991} = 0.0838. \tag{16}$$

where $\hbar = \frac{h}{2 \cdot \pi}$ is the reduced Planck constant, and $\omega_{ln} = exp\left[\frac{\int_0^\infty \frac{ln(\omega)}{\omega} \cdot F(\omega) \cdot d\omega}{\int_0^\infty \frac{1}{\omega} \cdot F(\omega) \cdot d\omega}\right]$, where $F(\omega)$ is the phonon density of states.

In result, the plot of $\frac{2 \cdot \Delta(0)}{k_B \cdot T_C}$ vs $\frac{k_B \cdot T_C}{\hbar \cdot \omega_{ln}}$ (which is often considered as an universal plot for phonon-mediated superconductors [72-75]), *C2/m*-SnH$_{12}$ phase falls into the lower branch (Fig. 5), where its NRTS contemplate H$_3$S is located [76].

It should be stressed, that in Fig. 5,a both fitting curves (red and cyan) and their 95% confidence band were not altered from ones in Fig. 4 in Ref. 76, because new fits were not performed (more details about these branches can be found in Ref. 76). It can be seen an unprecedented accuracy for the positioning of *C2/m*-SnH$_{12}$ phase in the lower branch. It should be noted that data on the upper branch in Fig. 5 with a very high accuracy can be described by simple elegant equation (Eq. 24 in Ref. 76):

$$\frac{2 \cdot \Delta(0)}{k_B \cdot T_c} = 3.53 \cdot \left(1 + 3.53 \cdot \left(\frac{k_B \cdot T_c}{\hbar \cdot \omega_{ln}}\right)^{1.29}\right) \tag{17}$$

In Fig. 5,b we fit data for lower branch (i.e. for $Pb_{0.5}Bi_{0.5}$, $Pb_{0.75}Bi_{0.25}$, Ga, Bi, $H_3S$ and $C2/m$-$SnH_{12}$) to equation [76]:

$$\frac{2 \cdot \Delta(0)}{k_B \cdot T_c} = A \cdot \left(1 + 3.53 \cdot \left(\frac{k_B \cdot T_c}{\hbar \cdot \omega_{ln}}\right)^{1.29}\right) \tag{18}$$

where A is free fitting parameter. It can be seen that 95% confidence band becomes narrower in Fig. 5,b in comparison with Fig. 5,a. Deduced parameter A = 2.86 ± 0.05 is practically undistinguishable from deduced A = 2.87 ± 0.06 reported in Ref. 76 for this parameter.

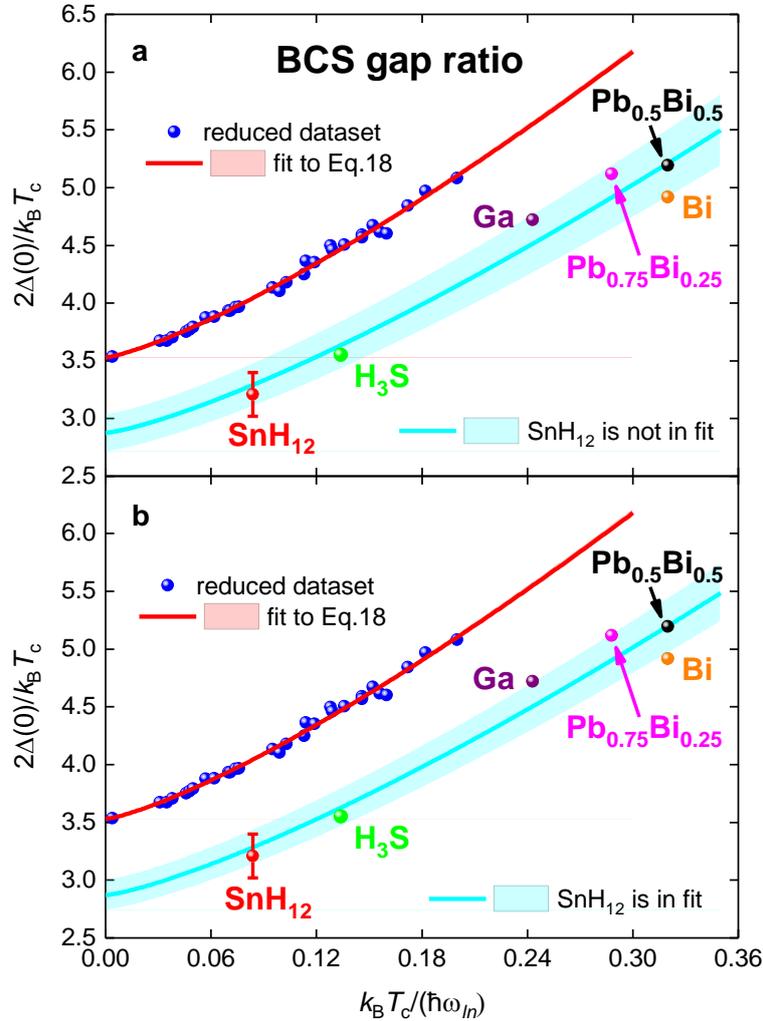

**Figure 5.** Full dataset of $\frac{2 \cdot \Delta(0)}{k_B \cdot T_c}$ vs $\frac{k_B \cdot T_c}{\hbar \cdot \omega_{ln}}$ from Table IV of Ref. 74 and data points for highly-compressed $H_3S$ and $SnH_{12}$. Fits to Eq. 17 (blue data points, red curve) and Eq. 18 (cyan curve) are shown. **a** - $SnH_{12}$ does not include in the fit (the fit is a clone from one in Fig. 4 of Ref. 59). **b** - $SnH_{12}$ does include in the fit. $A = 2.86 \pm 0.05$ and $R = 0.948$. 95% confidence bars are shown by a cyan shaded area.

**IV. *C2/m*-SnH$_{12}$ in the Uemura plot**

Uemura *et al* [31,32] reported empirical discovery that all unconventional superconductors, i.e. heavy fermions, cuprates, fullerenes and, later, to this list were added the iron-based superconductors [76,78] and hydrogen-rich superconductors [42,79-81], have the ratio of the superconducting transition temperature, $T_c$, to the Fermi temperature, $T_F$, within a narrow range:

$$0.01 \lesssim \frac{T_c}{T_F} \lesssim 0.05, \qquad (19)$$

while conventional superconductors have much smaller $\frac{T_c}{T_F}$ ratio:

$$\frac{T_c}{T_F} \lesssim 0.001 \qquad (20)$$

It should be noted that maximal value of $\frac{T_c}{T_F} = 0.22$ is attributed Bose-Einstein condensates (BEC). Thus, further step to characterize the superconducting state in *C2/m*-SnH$_{12}$ phase ($P = 190$ GPa) is to find the $\frac{T_c}{T_F}$ ratio for this compound.

The Fermi temperature can be calculated by an equation [76]:

$$T_F = \frac{\pi^2}{8 \cdot k_B} \cdot \left(1 + \lambda_{e-ph}\right) \cdot \xi^2(0) \cdot \left(\frac{\alpha \cdot k_B \cdot T_c}{\hbar}\right)^2, \qquad (21)$$

where $\alpha = \frac{2 \cdot \Delta(0)}{k_B \cdot T_c}$, and $\lambda_{e-ph}$ is the electron-phonon coupling constant. For calculations we utilized $\lambda_{e-ph} = 1.25$ reported by Esfahani *et al* [29] who computed by first-principles calculations several parameters for *C2/m*-SnH$_{12}$ phase. The rest of parameters in Eq. 21, i.e. α, $T_c$, $\xi(0)$, we deduced from the analysis of $B_{c2}(T)$ data above.

In a result, calculated Fermi temperature is $T_F = 5{,}658 \pm 906\ K$, and in the Uemura plot (Fig. 6), *C2/m*-SnH$_{12}$ phase falls into unconventional superconductors band in a close proximity to YBa$_2$Cu$_3$O$_{7-\delta}$ cuprates and in the same $T_c/T_F$ band where all NRTS counterparts are located. To date, an understanding that NRTS materials exhibit unconventional superconductivity is becoming more acknowledged [18,83,84], because if the superconducting transition temperature, $T_c$, in hydrogen-rich compounds was reasonably well predicted in some (and, what is important to stress, not in all) hydrogen-rich compounds, other calculated superconducting parameters, in particular, the ground state upper critical field and the ground state London penetration depth, are different from experimental values in several times.

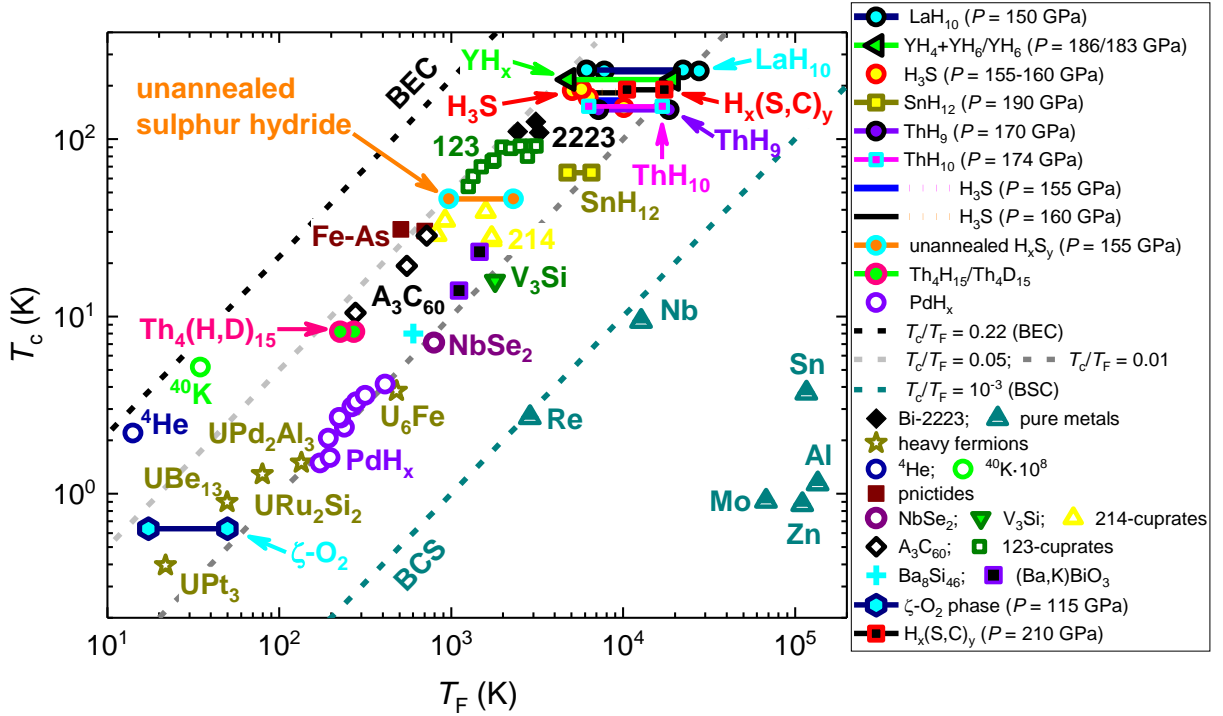

**Figure 6.** $T_c$ vs $T_F$ plot where the $C2/m$-$SnH_{12}$ ($P$ = 190 GPa) phase is shown together with main superconducting families: elemental superconductors, heavy-fermions, pnictides, cuprates, and near-room-temperature superconductors. Reference on original data can be found in Refs. 31,32,42,77-82. Boundary lines for BCS superconductors, for Bose-Einstein condensates and for $T_c/T_F$ = 0.05, 0.01 are shown.

### V. Conclusions

Recently, Hong *et al* [28] discovered a new highly-compressed $C2/m$-$SnH_{12}$ superhydride phase which exhibits the superconducting transition temperature of $T_c$ = 70 K at pressure of 190 GPa. Here we analyse the magnetoresistance data in this phase and deduce the ground state superconducting gap of $\Delta(0)$ = 9.15 ± 0.51 meV and the ratio of $2\Delta(0)/k_B T_c$ = 3.28 ± 0.18. Taking in account results of first principles calculations for this phase performed by Esfahani *et al* [29], we calculate the Fermi temperature $T_F = 5{,}658 \pm 906\ K$ in this phase, which means that in the Uemura plot [31,32], this new superhydride falls to unconventional superconductors band, where all other hydrogen-rich counterparts, including near-room-temperature superconductors, are located.

**Acknowledgement**

The author thanks financial support provided by the Ministry of Science and Higher Education of Russia (theme "Pressure" No. AAAA-A18-118020190104-3) and by Act 211 Government of the Russian Federation, contract No. 02.A03.21.0006.